\newcommand*{\ATLASLATEXPATH}{}
\pdfoutput=1 
\documentclass[cernpreprint, USenglish, texlive=2017]{\ATLASLATEXPATH atlasdoc}
\usepackage[subcaption]{\ATLASLATEXPATH atlaspackage}
\usepackage{\ATLASLATEXPATH atlasbiblatex}
\usepackage{\ATLASLATEXPATH atlascontribute}
\usepackage{\ATLASLATEXPATH atlasphysics}
\usepackage{subcaption}
\usepackage{siunitx}

\newcommand{\macro}[1]{\texttt{\textbackslash#1}}
\newcommand{\m}[1]{\macro{#1}}

\DeclareOldFontCommand{\rm}{\normalfont\rmfamily}{\mathrm}
\AtlasTitle{Search for the production of a long-lived neutral particle decaying within the ATLAS hadronic calorimeter in association with a $Z$ boson from $pp$ collisions at $\sqrt{s}=13\,\tev$}
\AtlasRefCode{EXOT-2017-24}
\PreprintIdNumber{CERN-EP-2018-275}
\AtlasJournal{Phys.\ Rev.\ Lett.}
\AtlasDOI{https://doi.org/10.1103/PhysRevLett.122.151801}
\AtlasJournalRef{PhysRevLett.122.151801}
\AtlasAbstract{%
	This Letter presents a search for the production  of a long-lived neutral particle ($Z_d$) decaying within the ATLAS hadronic calorimeter, 
	in association with a Standard Model (SM) $Z$ boson produced via an intermediate scalar 
	boson, where  $Z\to \ell^+\ell^-$ ($\ell=e,\mu$). 
	The data used were collected by the ATLAS detector during 2015 and 2016 \pp collisions with a center-of-mass energy of 
	$\sqrt{s} =$ 13\,\tev\ at the Large Hadron Collider and correspond to an integrated luminosity of $36.1\pm0.8$~fb$^{-1}$. 
	No significant excess of events is observed above the expected background. Limits on the production cross section of the scalar boson times its decay branching fraction into the long-lived neutral particle are derived as a function of the mass of the intermediate scalar boson, the mass of the long-lived neutral particle, and its 
	$c\tau$ from a few centimeters to one hundred meters.  
	In the case that the intermediate scalar boson is the SM Higgs boson, its decay branching fraction to a long-lived neutral particle with a $c\tau$ approximately between 0.1 and 7\,m is excluded with a 95\% confidence level up to 10\% for $m_{Z_d}$ between 5 and 15\,\gev.
}
\AtlasCoverSupportingNote{Support note}{https://cds.cern.ch/record/2267955}
\AtlasCoverEgroupEditors{atlas-EXOT-2017-24-editors@cern.ch}
\AtlasCoverEgroupEdBoard{atlas-EXOT-2017-24-editorial-board@cern.ch}

\hypersetup{pdftitle={ATLAS document},pdfauthor={The ATLAS Collaboration}}

\bibliography{ANA_EXOT-2017-24.bib}
\graphicspath{{logos/}{figures/}}

\begin{document}

\maketitle

Many extensions to the Standard Model (SM) such as supersymmetry~\cite{Dimopoulos:1996vz,Barbier:2004ez}, inelastic dark matter~\cite{TuckerSmith:2001hy}
and  hidden valley scenarios~\cite{Strassler:2006im,Strassler:2006ri} predict the existence of long-lived neutral particles that can decay hadronically. 
Search for long-lived neutral particles is an emerging field of research that has attracted significant theoretical and experimental interests. 
So far, only searches for the pair production of such particles have been carried out by the ATLAS~\cite{EXOT-2012-28,EXOT-2013-12,SUSY-2014-02,SUSY-2016-08}, 
CMS~\cite{CMS-EXO-12-038,CMS-EXO-16-003}, and  LHCb~\cite{Aaij:2014nma,Aaij:2016xmb}
experiments at the Large Hadron Collider (LHC), and the CDF~\cite{Aaltonen:2011rja} and D0~\cite{Abazov:2009ik} experiments at the Tevatron. 

This Letter reports a new way to look for new physics (NP) beyond the SM in a collider using singly produced long-lived neutral particle, which is one potential scenario  that NP
can manifest itself but had never been considered in theories or experiments. Among many possible single production final states, this Letter focuses on 
search for a hadronically decaying long-lived neutral particle, denoted by $Z_d$ hereafter, produced in association with a SM $Z$ boson through an intermediate scalar $\Phi$ or Higgs boson, $pp\rightarrow\Phi/H\rightarrow ZZ_d$, where $Z\to \ell^+\ell^-$ ($\ell=e,\mu$). 
Production of a new particle in association with a $Z$ boson is a popular scenario in hidden- or dark-sector models with an additional $U(1)_d$ dark gauge symmetry~\cite{Davoudiasl:2012ag,Davoudiasl:2013aya}. One such model has been tested by the ATLAS experiment in a search for a new particle that is mediated by the Higgs boson and decays promptly to a lepton pair~\cite{EXOT-2013-15, EXOT-2016-22}.
This analysis expands the search to a more general case to include a possible new scalar ($\Phi$) that couples to $Z$ and $Z_d$, instead of only the Higgs boson, and considers the scenario in which the $Z_d$ decays hadronically with a $c\tau$ between a few centimeters and one hundred meters, where $c$ is the speed of light and $\tau$ is the $Z_d$ proper lifetime.

The analysis uses data from $\sqrt{s}=13\,\tev$ proton--proton ($pp$) collisions at the LHC that were recorded by the ATLAS detector in 2015 and 2016 with single-electron and single-muon triggers~\cite{TRIG-2016-01}, corresponding to an integrated luminosity of 36.1$\pm$0.8~\ifb. 
The ATLAS detector~\cite{PERF-2007-01} is a multipurpose particle detector with a cylindrical geometry
\footnote{ATLAS uses a right-handed coordinate system with its origin at the nominal interaction point (IP) in the center of the detector and the $z$--axis along the beam pipe. The $x$--axis points to the center of the LHC ring, and the $y$--axis points upward. Cylindrical coordinates $(r, \phi)$ are used in the transverse plane, $\phi$ being the azimuthal angle around the $z$--axis. The pseudorapidity is defined
in terms of the polar angle $\theta$ as $\eta = - \ln \tan (\theta/2)$. The distance between two objects in the $\eta$--$\phi$ space is $\Delta R = \sqrt{(\Delta \eta)^2 + (\Delta \phi)^2}$. Transverse momentum is defined by $p_{T} = p \sin \theta$.}.
It consists of an inner detector (ID)~\cite{Abbott:2018ikt} surrounded by a solenoid that produces a 2~T magnetic field, electromagnetic and hadronic calorimeters, and a muon spectrometer in a magnetic field produced by a system of toroid magnets. 
The ID measures the trajectories of charged particles over the full azimuthal angle and in a pseudorapidity
range of $|\eta| < 2.5$ using silicon pixel, silicon microstrip, and straw-tube transition-radiation tracker detectors. 
Liquid-argon electromagnetic calorimeters (LArCal) extend from $1.5$ to $2.0\,\mbox{m}$ in radius in the barrel  
and from $3.6$ to $4.25\,\mbox{m}$  in $|z|$ in the endcaps. A scintillator-tile calorimeter (TileCal) provides hadronic  
calorimetery and covers the region $2.25 < r < 4.25\,\mbox{m}$.   
The experimental signature searched for is the $Z_d$ decaying within the TileCal, thus producing a jet that has little or no energy deposited in the LArCal, and no charged tracks that point to the reconstructed location of the collision of interest (hereafter called the primary vertex).

Monte Carlo (MC) simulated events are used to optimize the event selection and to help validate the analysis.
Signal samples were generated using the \PYTHIAV{8.210}~\cite{Sjostrand:2014zea}  generator with the NNPDF23LO parton distribution functions (PDFs) \cite{Ball:2014uwa} and the A14 set of tuned parameters (A14 tune)~\cite{ATL-PHYS-PUB-2014-021}, with an assumption that 
the $Z_d$ decays only to the highest-mass heavy quark pair ($b\bar{b}$ or $c\bar{c}$) that is kinematically allowed.
Nine samples were produced with three different $Z_d$ masses for each of three $\Phi$ masses 
($m_{Z_d}=\{5,10,15\}, \{10, 50, 100\}$, and  $\{20, 100, 200\}$ for $m_\Phi=125, 250,$ and $ 500\,\gev$, respectively),  where $m_\Phi=125\,\gev$ 
corresponds to the SM Higgs boson.  
The $c\tau$ of the $Z_d$ is a free parameter in this model. For each mass hypothesis of $Z_d$ and $\Phi$, its $c\tau$ is chosen to maximize the probability for $Z_d$ to decay inside the TileCal, which is found to be around $20\%$ for all samples, as shown in Figure 1(a). 
The events were reweighted to produce samples with different $c\tau(Z_{d})$ \cite{SUSY-2014-02} between $0.01$ and $100\,\rm{m}$.
The dominant SM background arises from events with a $Z$ boson produced in association with jets ($Z$+jets),
where a jet mimics the experimental signature of $Z_d$ decay inside the TileCal due to the presence of long-lived SM particles ($K_L^0$, $\Lambda$, etc), 
out-of-time pileup (additional $pp$ collisions occurring in bunch-crossings just before and after the collision of interest), noise, detector inefficiencies, and beam-induced background.
Additional SM background processes include the production of top quarks and $W$+jets. The SM background MC samples are generated with the 
configurations described in Ref.~\cite{ATL-PHYS-PUB-2016-003} for $W/Z$+jets production, and Ref.~\cite{ATL-PHYS-PUB-2016-004} for $t\bar{t}$ and single top production.
The effect of multiple $pp$ interactions in the same and neighboring bunch crossings (pileup) is included by overlaying minimum-bias events simulated with \PYTHIAV{8.186} on each generated event in all samples. 
The generated samples were processed through a \GEANT4-based detector simulation~\cite{Agostinelli:2002hh,SOFT-2010-01} and the standard ATLAS reconstruction software. 

\begin{figure*}
	\begin{center}
		\includegraphics[width=0.31\textwidth]{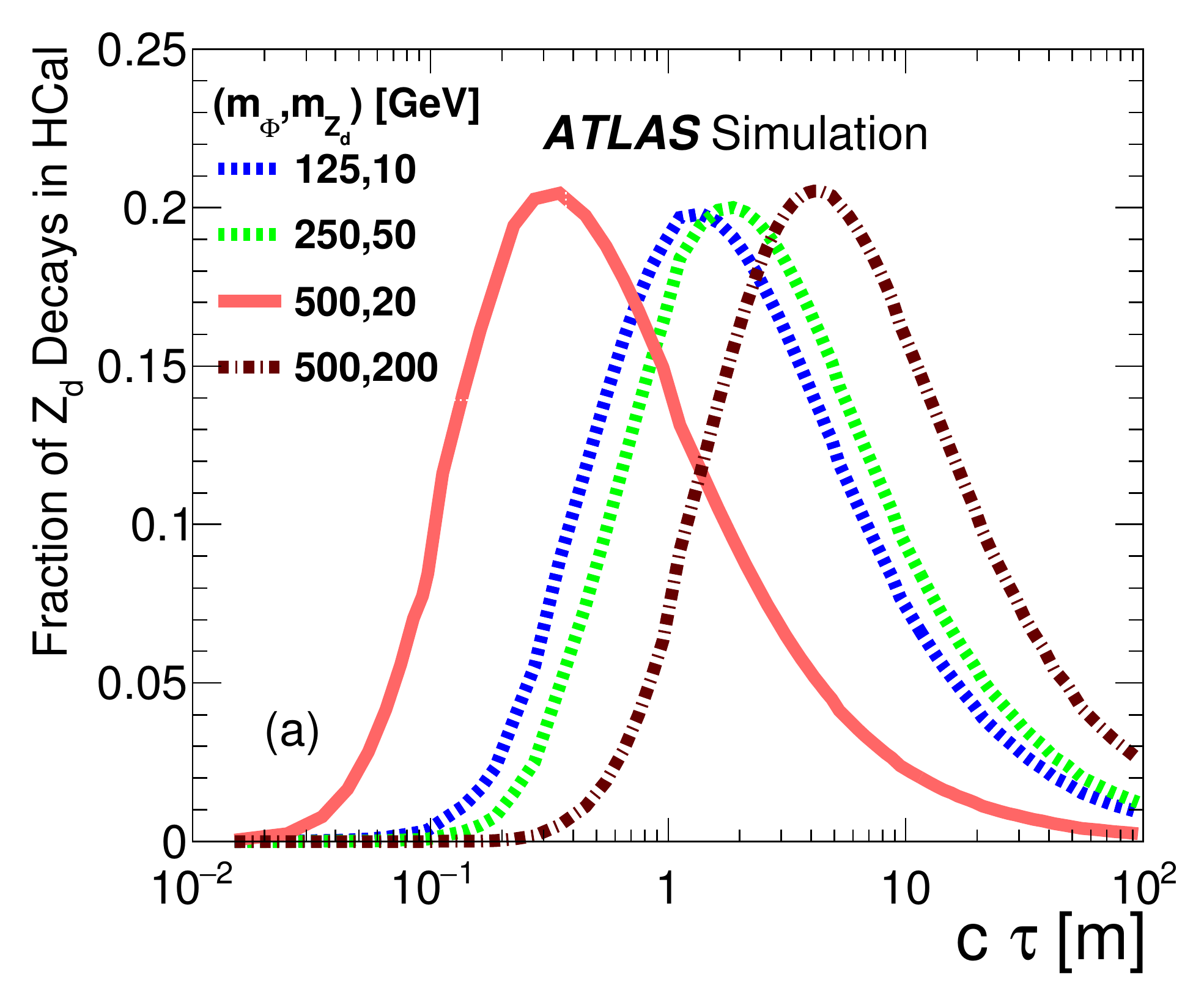}
		\includegraphics[width=0.31\textwidth]{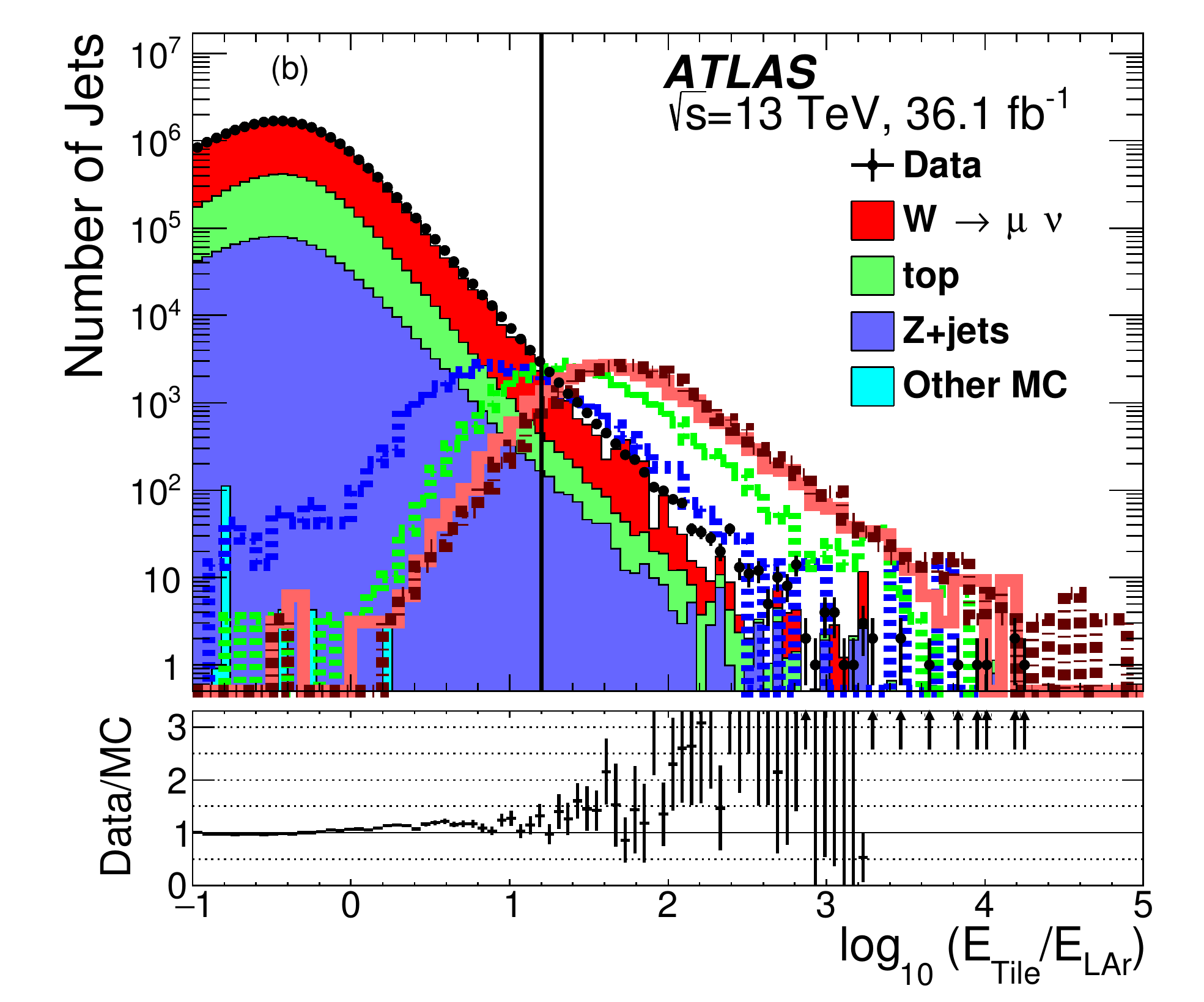}
		\includegraphics[width=0.31\textwidth]{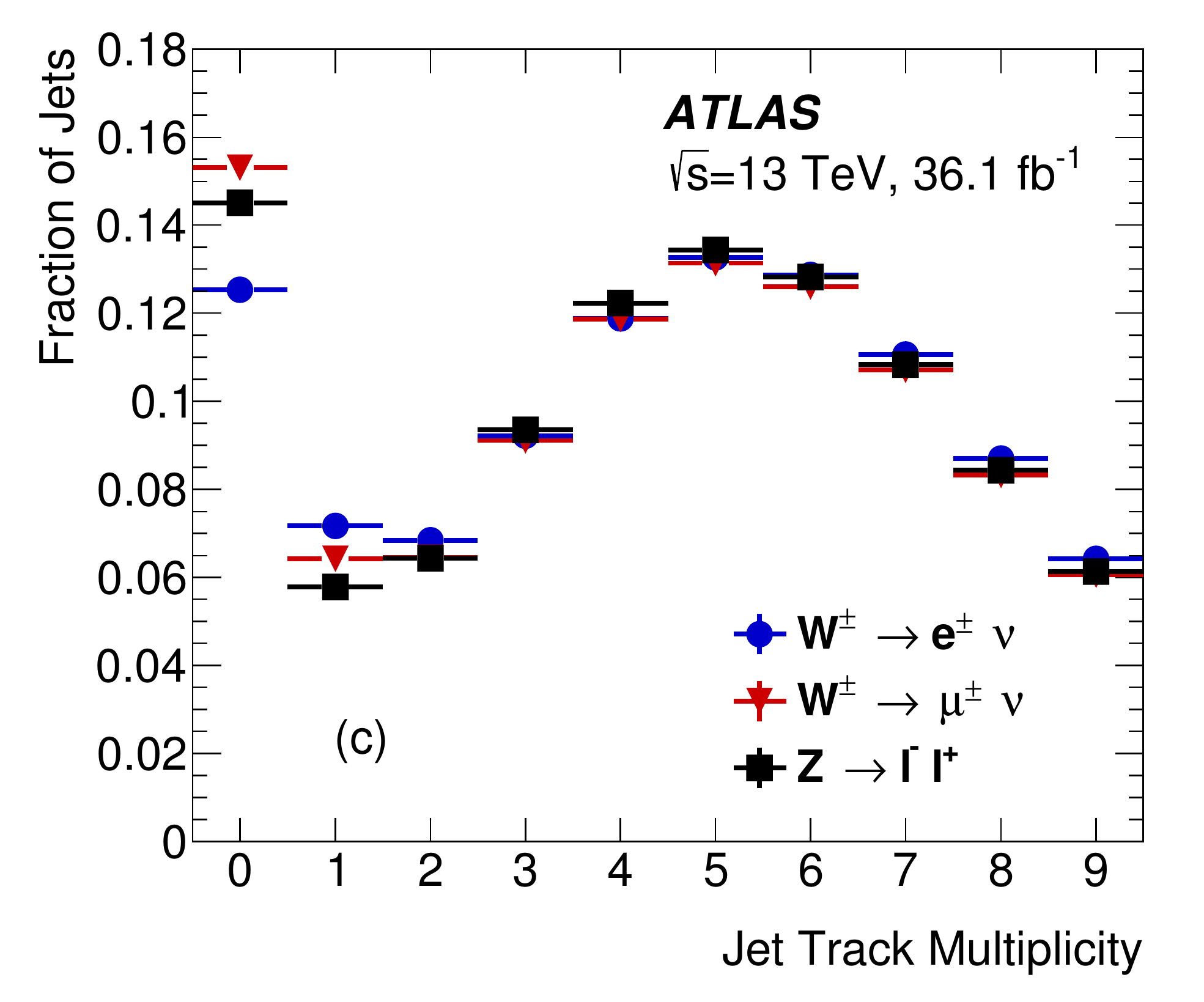}
		\caption{(a) The probability of a $Z_d$ boson to decay within the TileCal as a function of the $c\tau$ for each choice of $m_\Phi$ and $m_{Z_d}$. As $m_{Z_{d}}$ increases (for a fixed $m_{\Phi}$) the $Z_d$ becomes less boosted and therefore travels less distance into the detector before decaying.
		(b) The distributions of $\log_{10}({E_{\rm Tile}}/{E_{\rm LAr}})$ for jets in background and signal MC simulations (see legend of Figure 1(a) for signal labels) and $W$+jets data (prior to any requirements on the track multiplicity of jets or jet timing).  The threshold for this variable is shown as a solid black line. 
		(c) The distributions of the track multiplicity for jets prior to the selection of CR-jets in the $W$+jets and $Z$+jets data samples.
		}
		\label{fig:ZdCharastics}
	\end{center}	
\end{figure*}

The selected events have a pair of oppositely charged and isolated electrons~\cite{ATLAS-CONF-2016-024} 
or muons~\cite{PERF-2015-10} to form a $Z$ boson candidate. Electrons and muons are required to have $|\eta| <2.47$ 
and $|\eta|$ < 2.4, respectively, and $\pT > 25\,\gev\ (27\,\gev)$ in data collected in 2015 (2016).  The invariant mass of the $Z$ candidate ($m_{\ell\ell}$) 
is required to be between $66$ and $116\,\gev$. Selected jets must have transverse energy  $\et>40\,\gev$ and $|\eta|<2.0$ to ensure the jets are completely within the ID. They are reconstructed using the 
anti-$k_t$ algorithm~\cite{Cacciari:2008gp,Cacciari:2011ma} with a radius parameter $R=0.4$ and calibrated to particle level~\cite{PERF-2016-04}.
Standard ATLAS jet-quality criteria~\cite{ATLAS-CONF-2015-029} are applied, except the one for the ratio of the energy deposited in the hadronic calorimeter to the total energy since it removes signal jets.
A jet is considered as a $Z_d$ candidate, referred to as a calorimeter-ratio jet (CR-jet) hereafter, if it satisfies $\log_{10}({E_{\rm Tile}}/{E_{\rm LAr}}) > 1.2$ with no associated tracks \cite{Cacciari:2007fd} of $\pt>1\,\gev$ originating from the primary vertex, where $E_{\rm{Tile}}$ and $E_{\rm LAr}$ are the jet energy deposited in the TileCal and LArCal, respectively~\cite{EXOT-2012-28}, as shown in Figure 1(b).  
Jets with $\et < 60\, \gev\,$ in the transition region between the barrel and endcap cryostats ($1.0 < |\eta| < 1.3$) are not considered as CR-jet candidates due to noise in the gap scintillator of the TileCal~\cite{ATLAS-CONF-2016-042}.
In addition, the timing of the CR-jet is required to be between $-3\,\rm ns$ and $15\,\rm ns$ in order to suppress jets arising from out-of-time pileup and beam-induced backgrounds~\cite{EXOT-2012-28}. The timing of a jet is obtained from its constituent calorimeter cells by calculating an average time over cells weighted by cell-energy squared where the cell time is measured according to the bunch crossing clock, relative to the expected time-of-flight from the bunch crossing to the cell~\cite{PERF-2014-07}.
After this selection, the number of selected events containing a CR-jet with an \et\ above a chosen threshold is compared with the predicted total number of background events. 
The minimum \et\ requirement of the selected CR-jets is further optimized to achieve the highest experimental sensitivity for each mass hypothesis~\cite{Read:2002hq}.
It is set to be $40\,\gev$ for $m_\Phi = 125\,\gev$ samples, $60\,\gev$ for $m_\Phi = 250\,\gev$ samples, and $80\,\gev$ for $m_\Phi = 500\,\gev$ samples.

The signal efficiency times acceptance ($\epsilon\times A$) is defined as the ratio of the number of selected signal events in MC simulations to the number of generated signal events. 
It is a function of $m_\Phi$, $m_{Z_d}$ and the  $c\tau(Z_d)$. The maximum values vary between approximately 1\% for lowest $m_\Phi$ samples to 5--7\% for samples with larger $\Phi$ mass. 
The main loss is due to the low probability that $Z_d$ decays inside the TileCal, as shown in Figure 1(a).
The samples for $m_{\Phi}=125\,\gev$ suffer further efficiency loss due to the jet \et\ requirement.

MC simulations are not reliable enough to estimate the backgrounds of this analysis, as illustrated by the right hand side of Figure 1(b). 
A data-driven approach is thus used for its estimation.
A control data sample of SM $W$+jets events with the same event selection criteria of $W\to\ell\nu\;(\ell=e,\mu)$ in Ref.~\cite{STDM-2015-03}, is used to derive the probability for a jet to pass the selection of the CR-jet, assuming that the $Z_d$ cannot be produced in association with a $W$ boson.
The probability is calculated as $f_{\rm CR}=N_{\textrm{CR-jet}}/N_{\rm jet}$ in bins of the jet \et\ and $\eta$, 
where $N_{\textrm{CR-jet}}$ is the number of jets that satisfy the CR-jet selection criteria and $N_{\rm jet}$ is the total number of jets from the $W$+jets sample in each bin, as summarized in Table~\ref{Table:CRJetTotals}. 
For a selected event in data containing a $\Zboson\to\ell\ell$ candidate and $N$ jets, the corresponding probability for it to be identified as a signal event is therefore
$P= 1 - \prod_{i \in N} (1 - f_{\mathrm{CR}} (\et^i, \eta^i))$,
where $f_{\mathrm{CR}}(\et^i, \eta^i)$ is the probability of the $i$-th jet in the event to satisfy the CR-jet selection criteria.
The sum of the probabilities $P$ for all the selected events is therefore the expected number of background events.
Potential signal contamination of this control region was estimated using MC and found to have a $<1\%$ impact on the background estimate.

\begin{table}[!htb]
	\begin{center}
         \caption{The numbers of jets satisfying different requirements on minimum jet \et\ and their corresponding averaged CR-jet selection probabilities 			
         in the \Wboson \ra \lnu samples.}	
         \vspace{5pt}	
		\begin{tabular}{| l | r | r | r |}\hline
			Minimum jet $\et$ & $40\,\gev$   & $60\,\gev$    & $80\,\gev$    \\ \hline
			
			$N_{\textrm{CR-jet}}(W \rightarrow e \nu$) & $982$              & $189$                & $63$                \\
			$N_{\textrm{CR-jet}}(W \rightarrow \mu \nu$) & $1030$            & $186$                & $71$                \\
			\midrule
			$N_{\rm jet}(W \rightarrow e \nu$) & \multicolumn{1}{l|}{$3.3 \times 10^{7}$} & \multicolumn{1}{l|}{$1.5 \times 10^{7}$} & \multicolumn{1}{l|}{$0.8 \times 10^{7}$} \\
			$N_{\rm jet}(W \rightarrow \mu \nu$) & \multicolumn{1}{l|}{$3.1 \times 10^{7}$} & \multicolumn{1}{l|}{$1.3 \times 10^{7}$} & \multicolumn{1}{l|}{$0.7 \times 10^{7}$} \\
			\midrule
			$f_{\rm CR}(W \rightarrow e \nu$) & \multicolumn{1}{l|}{$3.0 \times 10^{-5}$} & \multicolumn{1}{l|}{$1.3 \times 10^{-5}$} & \multicolumn{1}{l|}{$7.9 \times 10^{-6}$} \\
			$f_{\rm CR}(W \rightarrow \mu \nu$) & \multicolumn{1}{l|}{$3.3 \times 10^{-5}$} & \multicolumn{1}{l|}{$1.4 \times 10^{-5}$} & \multicolumn{1}{l|}{$9.7 \times 10^{-6}$} \\
			\hline
		\end{tabular}
		\label{Table:CRJetTotals}
	\end{center}
\end{table}

Studies~\cite{EXOT-2012-28} have shown that jets originating from quarks and gluons may have different probabilities of satisfying the selection criteria for CR-jets.
MC simulations predict that jets from $W$+jets and $Z$+jets production are mostly initiated by quarks with a similar fraction ($\sim73\%$).
However, $W$+jets data samples are contaminated with a significant fraction of SM multijet events with a misidentified lepton,
which is estimated to be approximately $2\%$ in the muon final state and  $17\%$ in the electron final state using background-enriched control samples~\cite{STDM-2015-03}.
SM multijets originate primarily from gluons and thus introduce a difference between the $W$+jets and $Z$+jets samples.
The distributions of the track multiplicity of a jet in the $W/Z$+jets samples, which are sensitive to the quark/gluon jet fraction~\cite{ATL-PHYS-PUB-2017-009}, show a significant difference for track multiplicities of 0 and 1 in Figure 1(c). 
As a result, the $f_{\rm CR}$ values measured in the muon final state are used for the central value of the background estimate,
while the $f_{\rm CR}$ values measured in the electron final state are used as a cross-check to assign a systematic uncertainty due to different quark/gluon jet fractions in 
the $W$+jets and $Z$+jets samples.
The measured probabilities, $f_{\rm CR}$, are found to be dependent on the jet multiplicity in the event. Studies show that this is caused by the presence of jets from pileup interactions which deposit additional energy in the LArCal, suppressing the signature of CR-jets. 
The jet multiplicity and pileup distributions of events in the $W$+jets sample are the same as those from the $Z$+jets sample, and therefore the parameterization of the measured $f_{\rm CR}$ as a function of jet multiplicity or pileup is not necessary. 

Several studies were performed to validate the background estimation procedure. A $Z$+jets sideband is formed from events satisfying all signal selection criteria except the invariant-mass requirement for the \Zboson candidate. The mass is required to be $30 <m_{\ell\ell} < 55\,\gev$.
The events in the higher mass sideband $m_{\ell\ell}>116\,\gev$ are not used as they are still dominated by $Z$+jets production, as indicated by background MC simulations~\cite{STDM-2016-04}.
Based on the measured CR-jet probability in $W$+jets, the expected numbers of background events with \et of CR-jets greater than $40$, $60$, and $80\,\gev$ are estimated to be $2.2\pm 0.2$, $0.7\pm0.1$, and $0.3\pm0.1$, where the uncertainties are statistical only. 
They are consistent with the corresponding observations in data, which have $1$, $1$, and $0$ events, respectively.

The background estimation method relies on an assumption that jets in the $W$+jets sample have the same characteristics as jets in the $Z$+jets sample.
This assumption is tested using validation jets that are defined to satisfy the selection criteria of the CR-jets except the zero-ghost-track requirement.
Validation jets must have more than two associated tracks to avoid signal contamination, as MC-simulated signal events show that less than $1\%$ of jets from $Z_d$ decays inside the TileCal have more than two tracks. The probability for a jet to be identified as a validation jet is measured in the $W$+jets sample as a function of jet $\et$ and $\eta$ and subsequently used to predict the number of events containing a $Z\to\ell\ell$ candidate and at least one validation jet. 
As a result, a global scale factor of $1.24$, which is defined as the observed number of events with validation jets divided by the predicted value, is applied to the measured probabilities $f_{\rm CR}$. 
A $50\%$ relative correction of the scale factor ($\pm\,0.12$) is assigned as a systematic uncertainty due to potential bias of the background estimation procedure.

The systematic uncertainties of the background estimation include the statistical uncertainty from the $W$+jets sample (2--8\%), potential difference in the quark/gluon jet fractions between the $W/Z$+jets samples (7--20\%),  and the scale factor uncertainty ($\sim10\%$) measured using the validation jets. 
The uncertainty of the integrated luminosity is 2.1\%~\cite{DAPR-2013-01,Avoni:2018iuv}. 
Uncertainties resulting from detector effects such as the trigger efficiencies, the energy scale and resolution of jets~\cite{PERF-2016-04}, lepton identification, reconstruction and isolation efficiencies, lepton momentum scales and resolutions~\cite{ATLAS-CONF-2016-024,ATL-PHYS-PUB-2016-015,PERF-2015-10} only affect the calculation of the selection efficiencies of $Z_d$ signal events, since the background is estimated from the data. They are typically small (< 1--5\%). 
Pileup adds extra tracks and electromagnetic energy to jets.  The systematic uncertainties associated with reweighting the pileup distribution from the generated MC simulations to the data are typically small ($<5\%$) except for the samples with $m_\Phi=125\,\gev$ ($\sim13\%$), in which case the $Z_d$ have small energies and additional energy deposition in the LArCal from pileup can significantly affect their selection efficiencies.
Since the CR-jets in this analysis have a very small fraction of their energies inside the LArCal, the \textit{in situ} jet energy intercalibration~\cite{PERF-2016-04,EXOT-2012-28} is repeated using the $\pt$ balance method in dijets events, and the observed difference between the data and MC simulation is used to derive an additional systematic uncertainty of the jet energy scale.
The corresponding effect on the signal efficiencies is approximately 5--9\%  for samples with $m_\Phi=125\,\gev$, and negligible for samples with higher $m_\Phi$ values.   
The effects on the signal efficiency and acceptance due to theoretical uncertainties, such as PDF choice and initial- and final-state radiation modeling, are found to be very small ($<1\%$).

Table~\ref{Table:SignalResults} shows the predicted numbers of background events and the observed data events with different 
minimum \et\ requirements for the selected CR-jets. The data are well-described by the background estimate.
In the absence of any significant data excess, upper limits (ULs) on the signal yield of $pp\to \Phi \to Z Z_d$
at the 95\% confidence level (CL) are derived using the $\rm{CL}_s$ method~\cite{Read:2002hq} taking into
account both the statistical and systematic uncertainties. The results are listed in Table~\ref{Table:SignalResults}.
\begin{table}[!htb]
	\begin{center}
		\caption{Event yields for the predicted backgrounds and data, and the expected and observed ULs on the signal yields at the 95\% CL.
		The quoted errors include both the statistical and systematic uncertainties.}
		\vspace{5pt}
		\begin{tabular}{l | S[table-format=3.0] S[table-format=2.0] | S[table-format=2.1] S[table-format=1.1] | S[table-format=2.1] S[table-format=1.1] }
			\hline
			Minimum jet $\et$ & \multicolumn{2}{l|}{$40\,\gev$} & \multicolumn{2}{l|}{$60\,\gev$} & \multicolumn{2}{l}{$80\,\gev$} \\
			\hline
			Background  & 175 & \pm22 &  33.0  & \pm4.4 & 13.2 & \pm3.5 \\ 
			Data        & 158 &         &  35    &          & 16   &  \\ 
			\hline
			Expected UL &  65 &         &  17    &          & 10   & \\
			Observed UL &  50 &         &  18    &          & 13   & \\
			\hline
		\end{tabular}
		\label{Table:SignalResults}
	\end{center}
\end{table}

The results are further reinterpreted as the UL on the production cross section of $\Phi$ times the
decay branching fraction $B(\Phi\to ZZ_d)$,  as a function of $m_\Phi$, $m_{Z_d}$, and $c\tau$ 
of the $Z_d$. In the case of the SM Higgs boson, where $m_{H}=125\,\gev$, the UL on 
$B(H\to ZZ_d)$ are evaluated using the SM Higgs boson cross section $\sigma_{\rm SM}=48.5^{+4.6}_{-6.7}\,\rm pb$~\cite{deFlorian:2016spz}
of the gluon--gluon fusion process; other production modes are ignored. The results, reweighted to other $c\tau$ \cite{SUSY-2014-02}, are shown in Figure~\ref{fig:LimitPlotsBR}.

\begin{figure*}
\begin{center}
		\includegraphics[width=0.31\textwidth]{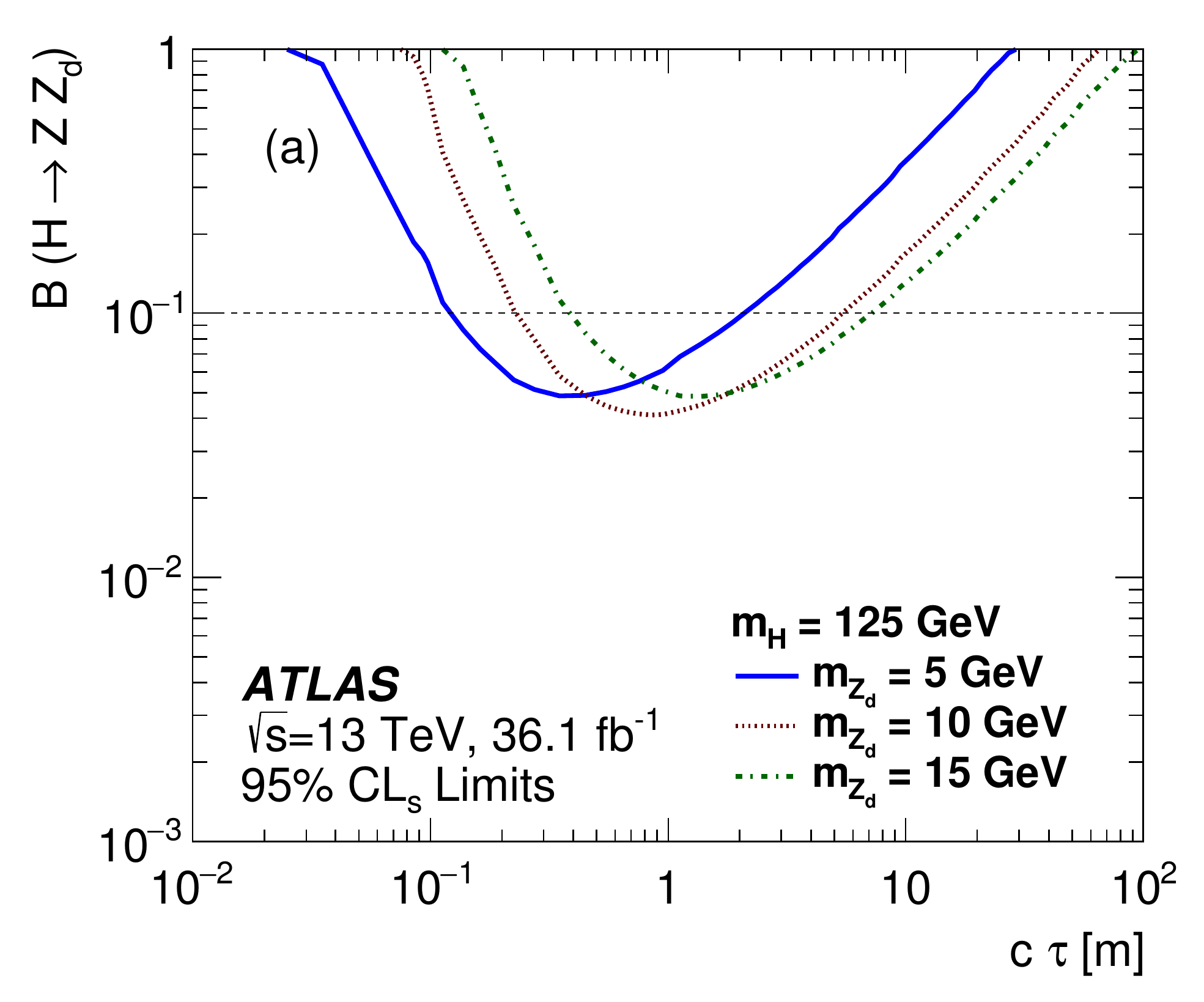}
		\includegraphics[width=0.31\textwidth]{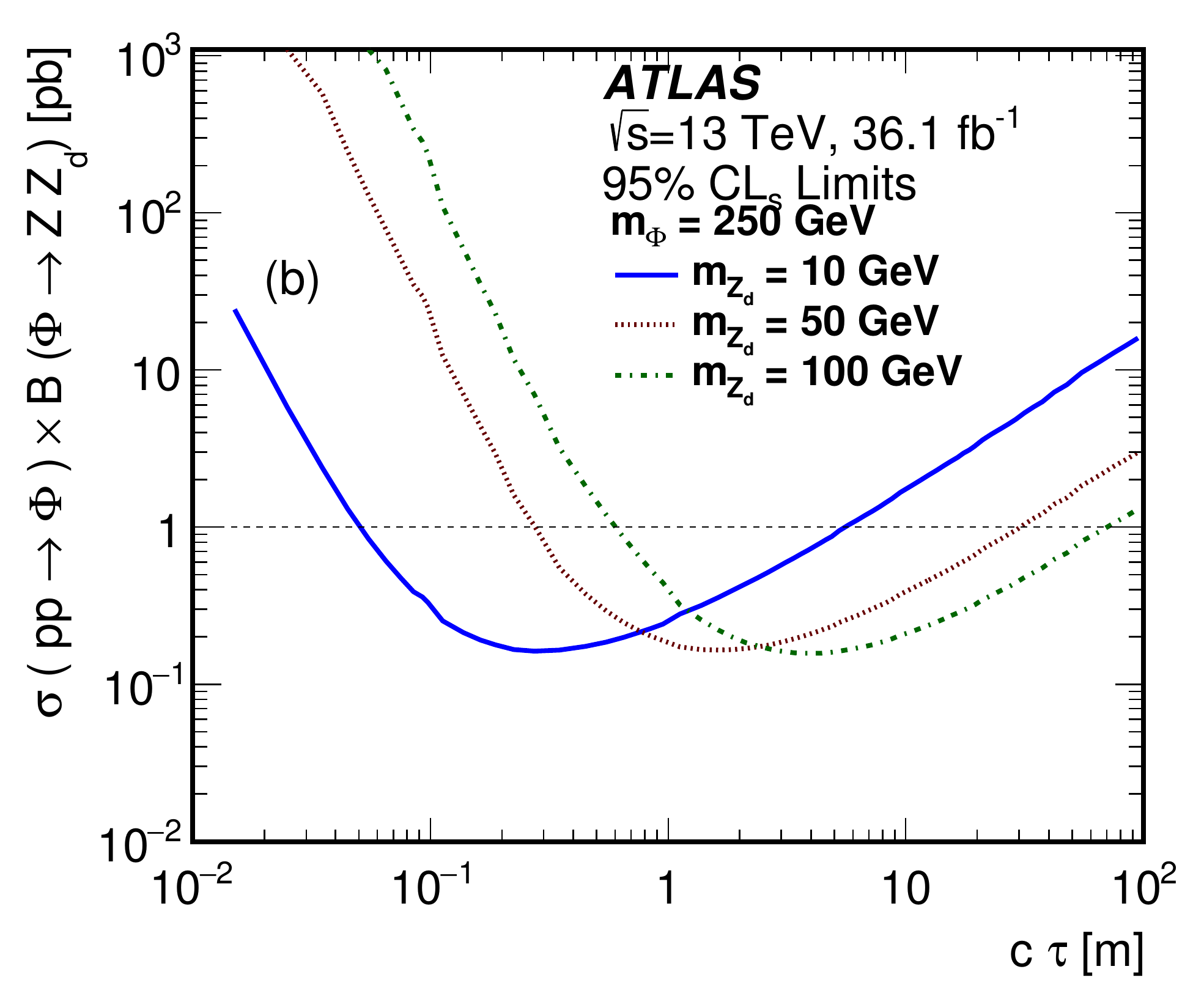}
		\includegraphics[width=0.31\textwidth]{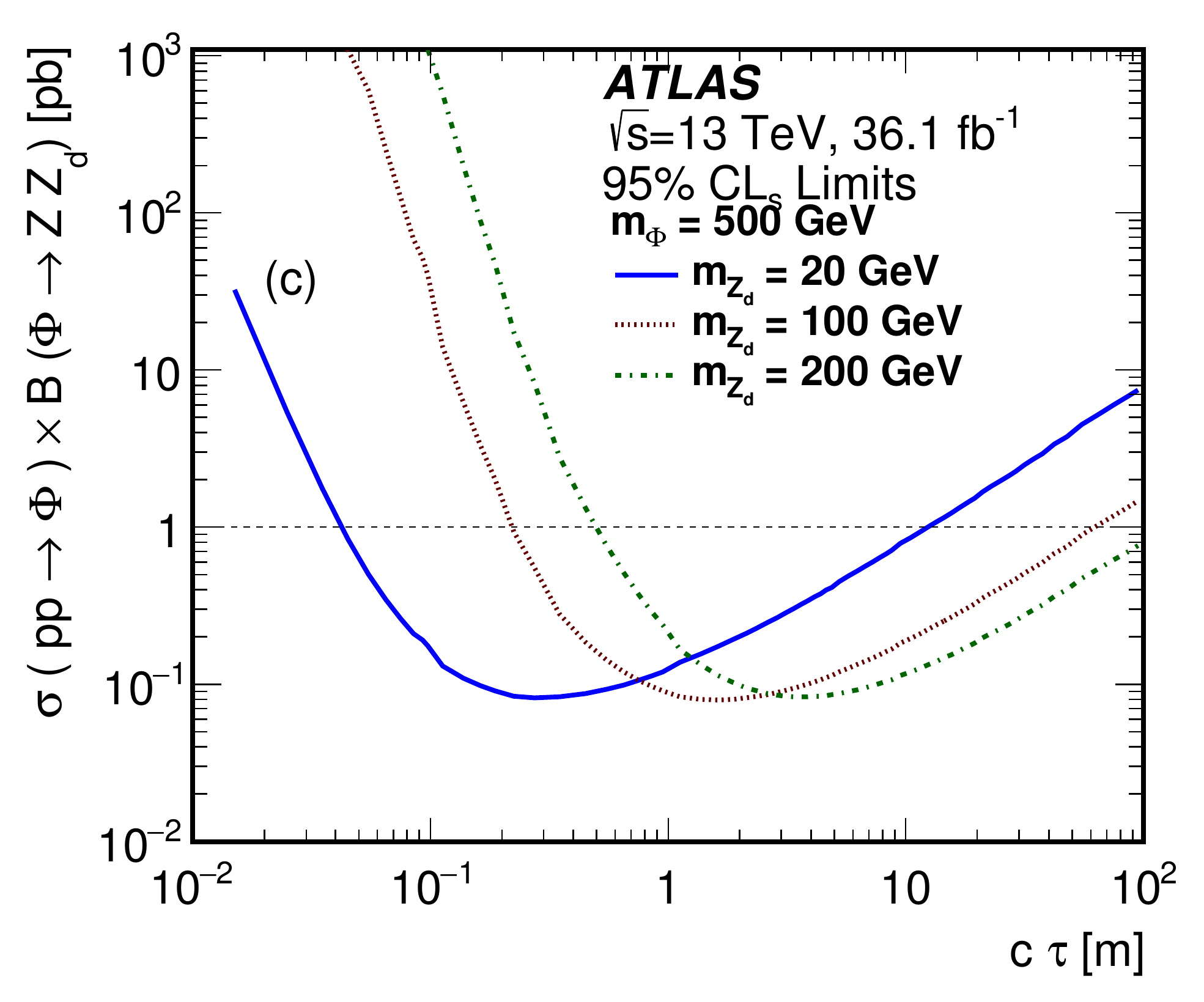}
		\caption{(a) Observed 95\% CL limits on the decay branching fraction of $B(H\to ZZ_d)$  for the SM Higgs boson as a function of the $c\tau(Z_d)$.  
		(b) and (c) Observed 95\% CL limits on the production cross section ($\sigma$) of $\Phi$ times its	decay branching fraction to $Z Z_d$ as a function of the $c\tau(Z_d)$. }
	\label{fig:LimitPlotsBR}
\end{center}	
\end{figure*}

In conclusion, this Letter reports a novel  search for a singly produced long-lived neutral particle $Z_d$, in association with an SM $Z$ boson via coupling to an intermediate scalar boson.
The analysis is based on $36.1 \pm 0.8\,\ifb$ of $pp$ collisions at $\sqrt{s}=13\,\TeV$ collected in 2015 and 2016 with the ATLAS detector at the LHC.
No excess over the expected background was observed.  Upper limits on the production cross section of the scalar boson times its branching fraction to the long-lived neutral particle 
at 95\% CL are derived as a function of the particle proper lifetimes for different masses of the scalar boson and the $Z_d$.  
In the case that the intermediate scalar boson is the SM Higgs boson, its decay branching fraction to a long-lived neutral particle with a $c\tau$
approximately between 0.1  and 7\,m is excluded with a 95\% CL up to 10\% for $m_{Z_d}$ between 5 and 15\,\gev.


We thank CERN for the very successful operation of the LHC, as well as the
support staff from our institutions without whom ATLAS could not be
operated efficiently.

We acknowledge the support of ANPCyT, Argentina; YerPhI, Armenia; ARC, Australia; BMWFW and FWF, Austria; ANAS, Azerbaijan; SSTC, Belarus; CNPq and FAPESP, Brazil; NSERC, NRC and CFI, Canada; CERN; CONICYT, Chile; CAS, MOST and NSFC, China; COLCIENCIAS, Colombia; MSMT CR, MPO CR and VSC CR, Czech Republic; DNRF and DNSRC, Denmark; IN2P3-CNRS, CEA-DRF/IRFU, France; SRNSFG, Georgia; BMBF, HGF, and MPG, Germany; GSRT, Greece; RGC, Hong Kong SAR, China; ISF and Benoziyo Center, Israel; INFN, Italy; MEXT and JSPS, Japan; CNRST, Morocco; NWO, Netherlands; RCN, Norway; MNiSW and NCN, Poland; FCT, Portugal; MNE/IFA, Romania; MES of Russia and NRC KI, Russian Federation; JINR; MESTD, Serbia; MSSR, Slovakia; ARRS and MIZ\v{S}, Slovenia; DST/NRF, South Africa; MINECO, Spain; SRC and Wallenberg Foundation, Sweden; SERI, SNSF and Cantons of Bern and Geneva, Switzerland; MOST, Taiwan; TAEK, Turkey; STFC, United Kingdom; DOE and NSF, United States of America. In addition, individual groups and members have received support from BCKDF, CANARIE, CRC and Compute Canada, Canada; COST, ERC, ERDF, Horizon 2020, and Marie Sk{\l}odowska-Curie Actions, European Union; Investissements d' Avenir Labex and Idex, ANR, France; DFG and AvH Foundation, Germany; Herakleitos, Thales and Aristeia programmes co-financed by EU-ESF and the Greek NSRF, Greece; BSF-NSF and GIF, Israel; CERCA Programme Generalitat de Catalunya, Spain; The Royal Society and Leverhulme Trust, United Kingdom. 

The crucial computing support from all WLCG partners is acknowledged gratefully, in particular from CERN, the ATLAS Tier-1 facilities at TRIUMF (Canada), NDGF (Denmark, Norway, Sweden), CC-IN2P3 (France), KIT/GridKA (Germany), INFN-CNAF (Italy), NL-T1 (Netherlands), PIC (Spain), ASGC (Taiwan), RAL (UK) and BNL (USA), the Tier-2 facilities worldwide and large non-WLCG resource providers. Major contributors of computing resources are listed in Ref.~\cite{ATL-GEN-PUB-2016-002}.

\printbibliography
\clearpage
 
\begin{flushleft}
{\Large The ATLAS Collaboration}

\bigskip

M.~Aaboud$^\textrm{\scriptsize 34d}$,    
G.~Aad$^\textrm{\scriptsize 99}$,    
B.~Abbott$^\textrm{\scriptsize 125}$,    
O.~Abdinov$^\textrm{\scriptsize 13,*}$,    
B.~Abeloos$^\textrm{\scriptsize 129}$,    
D.K.~Abhayasinghe$^\textrm{\scriptsize 91}$,    
S.H.~Abidi$^\textrm{\scriptsize 164}$,    
O.S.~AbouZeid$^\textrm{\scriptsize 39}$,    
N.L.~Abraham$^\textrm{\scriptsize 153}$,    
H.~Abramowicz$^\textrm{\scriptsize 158}$
% ...
\bigskip
\\

% ...
$^{13}$Institute of Physics, Azerbaijan Academy of Sciences, Baku; Azerbaijan.\\
% ...
$^{34}$$^{(a)}$Facult\'e des Sciences Ain Chock, R\'eseau Universitaire de Physique des Hautes Energies - Universit\'e Hassan II, Casablanca;$^{(b)}$Centre National de l'Energie des Sciences Techniques Nucleaires (CNESTEN), Rabat;$^{(c)}$Facult\'e des Sciences Semlalia, Universit\'e Cadi Ayyad, LPHEA-Marrakech;$^{(d)}$Facult\'e des Sciences, Universit\'e Mohamed Premier and LPTPM, Oujda;$^{(e)}$Facult\'e des sciences, Universit\'e Mohammed V, Rabat; Morocco.\\
% ...
$^{39}$Niels Bohr Institute, University of Copenhagen, Copenhagen; Denmark.\\
% ...
$^{91}$Department of Physics, Royal Holloway University of London, Egham; United Kingdom.\\
% ...
$^{99}$CPPM, Aix-Marseille Universit\'e, CNRS/IN2P3, Marseille; France.\\
% ...
$^{125}$Homer L. Dodge Department of Physics and Astronomy, University of Oklahoma, Norman OK; United States of America.\\
% ...
$^{129}$LAL, Universit\'e Paris-Sud, CNRS/IN2P3, Universit\'e Paris-Saclay, Orsay; France.\\
% ...
$^{153}$Department of Physics and Astronomy, University of Sussex, Brighton; United Kingdom.\\
% ...
$^{158}$Raymond and Beverly Sackler School of Physics and Astronomy, Tel Aviv University, Tel Aviv; Israel.\\
% ...
$^{164}$Department of Physics, University of Toronto, Toronto ON; Canada.\\
% ...
$^{a}$ Also at Borough of Manhattan Community College, City University of New York, NY; United States of America.\\
$^{b}$ Also at California State University, East Bay; United States of America.\\
$^{c}$ Also at Centre for High Performance Computing, CSIR Campus, Rosebank, Cape Town; South Africa.\\
$^{d}$ Also at CERN, Geneva; Switzerland.\\
$^{e}$ Also at CPPM, Aix-Marseille Universit\'e, CNRS/IN2P3, Marseille; France.\\
% ...
$^{*}$ Deceased

\end{flushleft}


\end{document}